\pdfoutput=1
\documentclass{JINST}

\title{High-speed data transfer with FPGAs and QSFP+ modules}

 \author{R. Ammendola$^a$, A. Biagioni$^b$, G. Chiodi$^b$, O. Frezza$^b$, F. Lo Cicero$^b$, A. Lonardo$^b$, R. Lunadei$^b$, P. S. Paolucci$^b$, D. Rossetti$^b$, A. Salamon$^a$\thanks{Corresponding author.}~, G. Salina$^a$, F. Simula$^b$, L. Tosoratto$^b$, P. Vicini$^b$\\
\llap{$^a$}INFN Sezione di Roma Tor Vergata,\\
  Via della Ricerca Scientifica, 1 - 00133 Roma Italy\\
\llap{$^b$}INFN Sezione di Roma\\
  P.le Aldo Moro, 2 - 00185 Roma Italy\\
  E-mail: \email{andrea.salamon@roma2.infn.it}}

\abstract{We present test results and characterization of a data transmission system based on a last generation FPGA and a commercial QSFP+ (Quad Small Form Pluggable +) module. QSFP+ standard defines a hot-pluggable transceiver available in copper or optical cable assemblies for an aggregated bandwidth of up to 40 Gbps. We implemented a complete testbench based on a commercial development card mounting an Altera Stratix IV FPGA with 24 serial transceivers at 8.5 Gbps, together with a custom mezzanine hosting three QSFP+ modules. We present test results and signal integrity measurements up to an aggregated bandwidth of 12 Gbps.}

\keywords{data acquisition circuits; digital electronic circuits}

\begin{document}

\section{Introduction}

High-speed data transfer plays a central role in trigger and data acquisition systems for present and future (sLHC, ILC,vCLIC, rare kaon decays) particle and astroparticle physics experiments.

We present test results and characterization of a data transmission system aimed at very high throughput applications based on last generation FPGA-embedded serial transceivers and commercial QSFP+ (Quad Small Form Pluggable +) modules.

This data transmission system was originally conceived for a commercial PC cluster for lattice QCD and other computing intensive numerical algorithms \cite{ape1},\cite{ape2},\cite{ape3}. However it can also be of interest for future particle and astroparticle data acquisition systems given its high bandwidth, low power and low cost characteristics.

\subsection{High-speed FPGA serial transceivers}

High-speed FPGA-embedded serial transceivers technology greatly evolved in the last 10 years and is now mature and reliable with multi Gbps transceivers easily available on commercial devices. In particular we focus on an Altera Stratix IV GX FPGA \cite{altera1} with embedded full duplex serial transceivers with a data rate from 600 Mbps to 8.5 Gbps, 8B10B encoding, clock and data recovery, channel bonding (up to 8x) and programmable pre-emphasis and equalization.

\subsection{QSFP+ standard}

QSFP+ (Quad Small Form Pluggable +) is an electrical and mechanical standard for point-to-point links over copper or optical fibers aimed at high-speed data rate, high-density and low-power applications with an aggregated bandwidth up to 40 Gbps per direction \cite{qsfp}. QSFP+ is the latest step in the evolution from SFP (Small Form Pluggable) standard. It defines an hot-pluggable transceiver with 4 transmit and 4 receive channels which physical layer is based on 100 Ohm AC coupled CML differential lines together with copper or fiber optic cables. Different interchangeable cable assemblies are now available on the market: passive copper cables for data transmission up to 5m, active copper (cable assembly with integrated amplifiers) for data transmission up to 15m and active optical modules for data transmission up to 100 meters.

\section{Test system}

 \begin{figure}
\centering
\includegraphics[width=.49\textwidth]{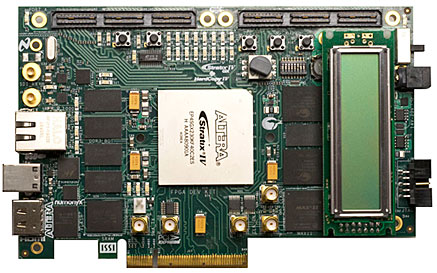}
\includegraphics[width=.49\textwidth]{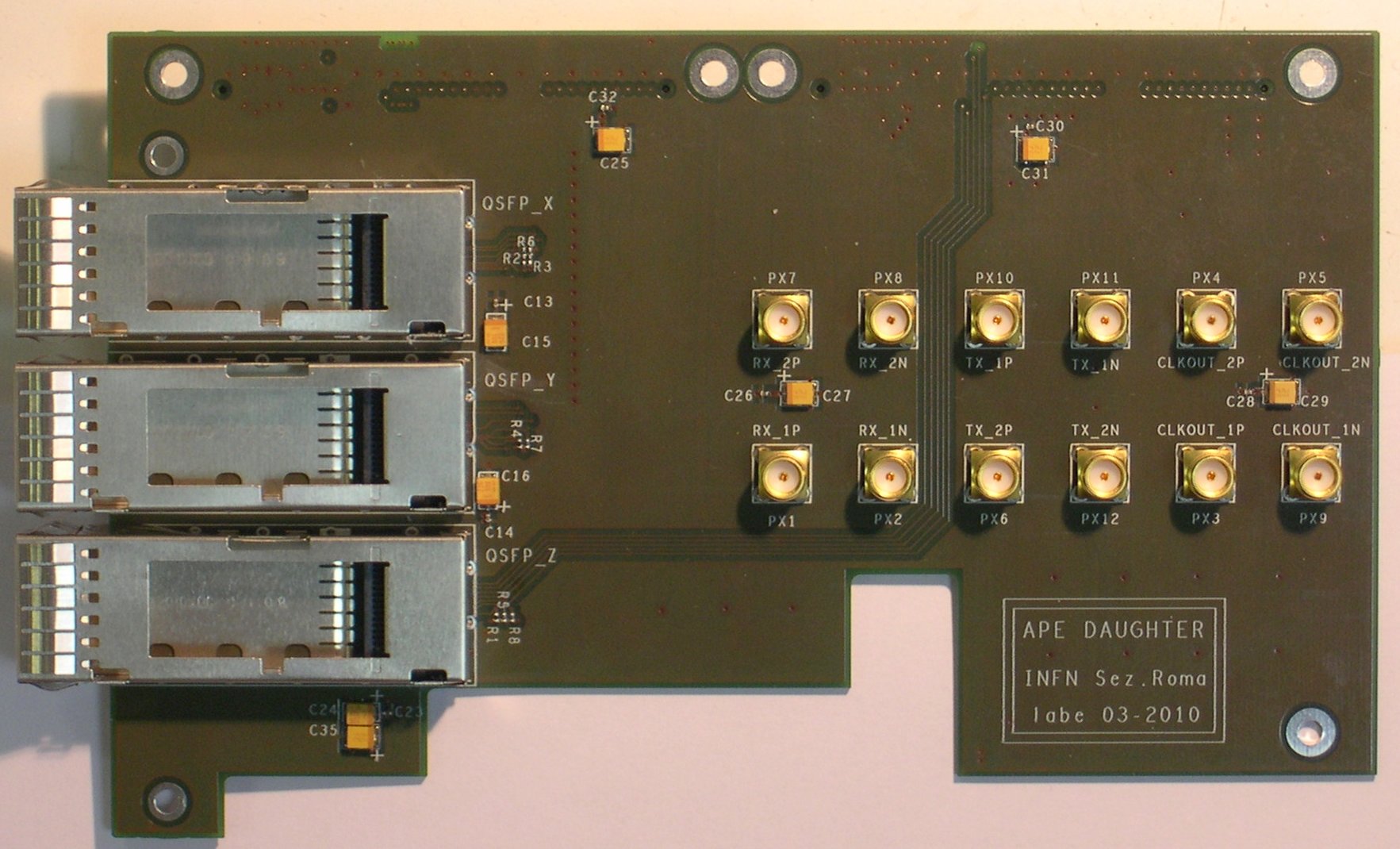}
 \caption{Altera Stratix IV GX 230 development kit (left) and custom PCI form factor mezzanine connecting the FPGA serial transceivers to the QSFP+ connectors (right).}
 \label{fig1}
 \end{figure}

The test system is composed of an Altera Stratix IV GX 230 development kit \cite{altera2}, based on an Altera Stratix IV FPGA with 230k Logic Elements, 14 Mbit of embedded memory and 24 serial transceivers (see figure \ref{fig1}). 8 serial transceivers are used for the PCI-Express gen 2 interface, 14 are routed to two Samtec high-speed connectors \cite{samtec1} and the remaining are connected to test points on the development kit. 

A PCI form factor custom mezzanine has been designed to be mounted on top of the Altera development kit (see figure \ref{fig1}). This mezzanine mounts two 19 mm high Samtec connectors \cite{samtec2}, three QSFP+ connectors and cage assemblies and some SMA test connectors. 12 of the 14 Altera high speed channels connected to the Samtec connectors are routed to three QSFP+ connectors while the remaining 2 are routed to SMA test points.

The test instrumentation is composed of a high stability few ps jitter clock generator, a 20 GHz sampling scope, an external clock data recovery module and a 500 MHz real time scope.
%The signal from the two differential lines are transmitted over two SMA cable to a high frequency passive splitter: half of the signal is sent to a clock data recovery module and the other half is sento to the sampling scope. {\bf bla bla bla}

\section{Test results}

The following tests were performed: signal integrity measurements, recovered clock, latency measurements and error rate measurements.

\begin{figure}
\centering
\includegraphics[width=.43\textwidth]{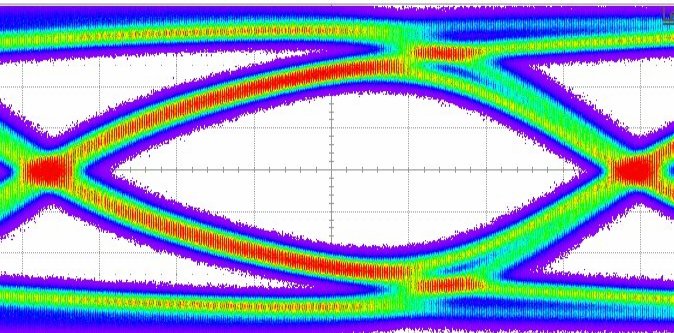}
\includegraphics[width=.44\textwidth]{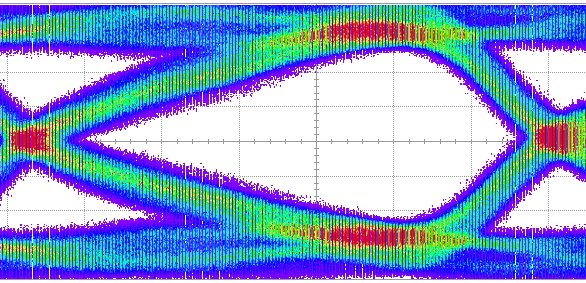}
 \caption{Eye diagram at 5 Gbps on the development kit (left) and eye diagram at 3 Gbps on the mezzanine card (right) after one Samtec connector, two QSFP+ connectors and 1 m of copper QSFP+ cable (right).}
 \label{fig2}
 \end{figure}

The signal integrity was checked with the high bandwidth sampling scope connected to dedicated SMA test points on the Altera demo board, immediately at the output of the FPGA transceivers (see figure \ref{fig2}). The signal integrity was also checked on the mezzanine card immediately after the Samtec high-speed tower connector and also after one meter of QSFP+ cable (see figure \ref{fig2}).

\begin{figure}
\centering
\includegraphics[width=.45\textwidth]{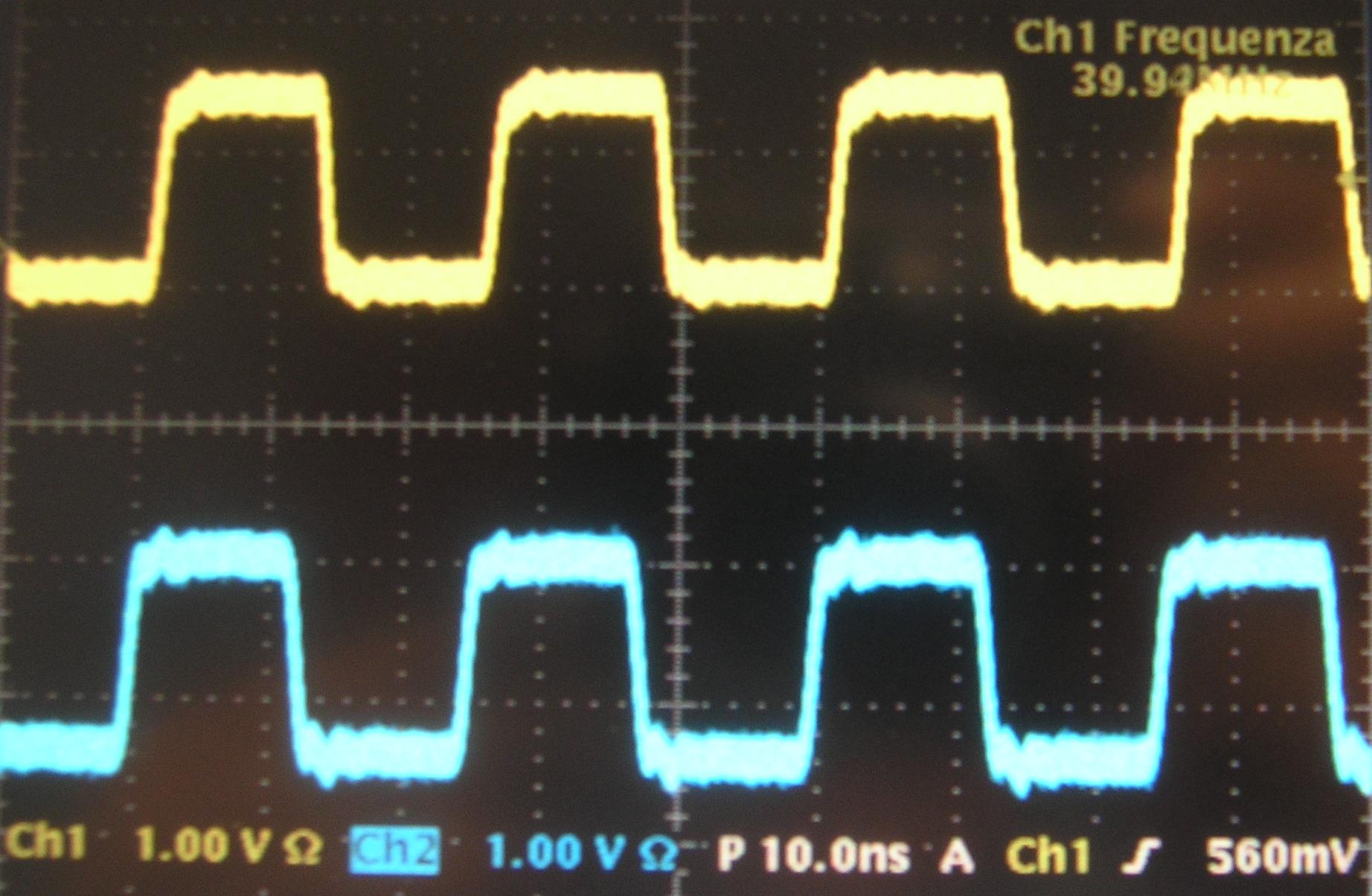}
\includegraphics[width=.43\textwidth]{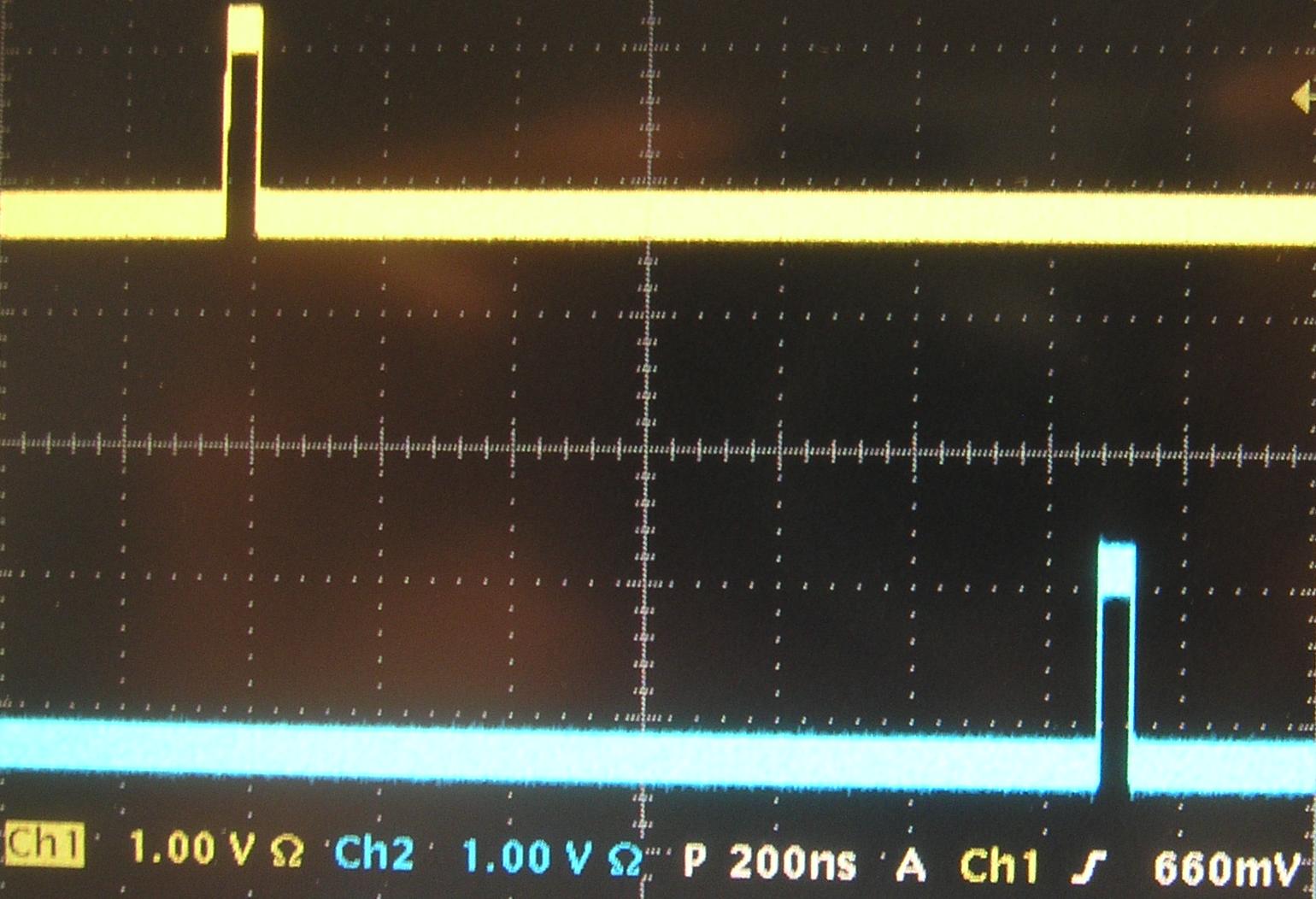}
 \caption{Recovered clock stability check at 40 MHz (left) and latency measurement at 40 MHz (right).}
 \label{fig3}
 \end{figure}

Recovered clock stability was checked transmitting a pseudorandom data stream organized in 128 bit wide words \footnote{Each one of the four serial transceivers grouped in a single QSFP+ link carries a 32 bit parallel word per input single data rate transceiver clock. The clock of transceiver on line 0 is used for channel bonding and as recovered clock for clock stability and latency measurements.} over 1 meter copper QSFP+ cable and checking the relative phase between the input and the output clocks (see figure \ref{fig3}). Recovered clock was found stable and in phase with the input clock up to 400 MHz.

Latency was checked transmitting a pseudorandom sequence over 1 meter QSFP+ copper cable and rising a flag every time a fixed test word is transmitted and received by the serializer and the deserializer respectively (see figure \ref{fig3}). Transmission system latency was found stable up to 160 MHz transmitting clock.

Long runs with single line data rate up to 3 Gbps (12 Gbps aggregated bandwidth on a single QSFP+ cable) were performed without any error.

The proposed link was successfully tested up to 3 Gbps data rate (compared to 8.5 Gbps achievable with the Stratix IV embedded transceivers). Above this limit we found some signal integrity degradation probably caused by the tower connector between the Altera development kit and the test mezzanine. Further investigation is still in progress, however this is perfectly reasonable given the reduced bandwidth of the 19 mm high QTH Samtec connector (below 5 GHz). We were forced to use this connector in our test mezzanine by the connector already mounted on the Altera development kit \footnote{Of course smaller and higher bandwidth Samtec QTH connectors are used by Altera to connect the TX and RX side of the transceivers with a small slice of PCB.} , higher bandwidth connectors will be used in the production release of the communication card for the Lattice QCD parallel processor \cite{samtec3}. 

Characterization of signal integrity (and maximum achievable bandwidth) versus serial trans\-cei\-vers pre-emphasis and equalization is still in progress, all presented measurements are still without any pre-emphasis or receiver equalization. Cables and optical fibers of different lengths will also be tested in the near future.

\section{Conclusions}

Test results and characterization of a data transmission system based on a last generation FPGA and a commercial QSFP+ module were presented. The link was successfully operated up to an aggregated bandwidth of 12 Gbps per QSFP+ cable. Further tests are planned with different lengths of cables and fiber optics and with the production release of the data transmission board mounting the high bandwidth connectors needed to exploit the full bandwidth of the Altera transceivers.
This data transmission system originally conceived for a commercial PC cluster for lattice QCD and other computing intensive numerical algorithm can also be of interest for future particle and astroparticle data acquisition systems.

\acknowledgments

The authors would like to thank the Electronics Laboratory at INFN Sezione di Roma \cite{labe} for technical support with the design, production and assembly of the test board used in this work.

This work was partially supported by the EU Framework Programme 7 project EURETILE under grant number 247846.


\begin{thebibliography}{9}

\bibitem{ape1}
\emph{http://apegate.roma1.infn.it/APE}

\bibitem{ape2}
R. Ammendola et al, \emph{Status of the APENet project}, in proceedings of
\emph{XXIIIrd International Symposium on Lattice Field Theory}, 25-30 July 2005 Trinity College, Dublin, Ireland
\pos{PoS(LAT2005)100}.

\bibitem{ape3}
R. Ammendola et al, \emph{APENet+: a 3D toroidal network enabling petaFLOPS scale Lattice QCD simulations on commodity clusters}, to appear in proceedings of
\emph{XXVIIIth International Symposium on Lattice Field Theory}, 14-19 June 2010 Villasimius, Sardinia, Italy

\bibitem{altera1}
\emph{http://www.altera.com/products/devices/stratixfpgas/stratix-iv/transceivers/stxiv-transceivers.html}

\bibitem{qsfp}
\emph{ftp://ftp.seagate.com/sff/SFF-8436.PDF}

\bibitem{altera2}
\emph{http://www.altera.com/products/devkits/altera/kit-siv-gx.html}

\bibitem{samtec1}
\emph{http://www.samtec.com/ProductInformation/TechnicalSpecifications/Overview.aspx?series=QSH}

% questa e' la torretta
\bibitem{samtec2}
\emph{http://www.samtec.com/ProductInformation/TechnicalSpecifications/Overview.aspx?series=QTH}

\bibitem{samtec3}
\emph{http://www.samtec.com/ProductInformation/TechnicalSpecifications/Overview.aspx?series=SEAF}

\bibitem{labe}
\emph{http://maclabe.roma1.infn.it}

\end{thebibliography}
\end{document}